\documentclass[journal, onecolumn,12pt, draftclsnofoot]{IEEEtran}

\ifCLASSINFOpdf
\else
   \usepackage[dvips]{graphicx}
\fi
\usepackage{url}
\usepackage{hhline}
\usepackage{graphicx}
\usepackage{multicol}
\usepackage{graphicx}
\usepackage{amssymb}
\usepackage{amsmath}
\DeclareMathOperator*{\argmax}{arg\,max}

\usepackage{calc}
\usepackage{array}
\usepackage{optidef}
\usepackage{epstopdf}
\usepackage{mathtools}
\usepackage{mathrsfs}
\usepackage{bm}
\usepackage{cite}
\usepackage{boldline}

\usepackage{booktabs,makecell,tabularx}

\newcolumntype{C}[1]{>{\centering\arraybackslash}p{#1}}
\newcolumntype{L}{>{\raggedright\arraybackslash}X}

\usepackage{siunitx}
\usepackage{etoolbox}
\newrobustcmd{\B}{\bfseries}

\usepackage{comment}
\usepackage{algorithm}
\usepackage{algpseudocode}
\usepackage{enumitem} 

\usepackage{colortbl}     
\usepackage{subfigure}
\usepackage{color, soul}
\usepackage{stfloats}
\usepackage{float}
\usepackage{multirow}
\usepackage{wrapfig}
\usepackage{booktabs}
\usepackage{slashbox}
\usepackage{optidef}
\definecolor{LightBlue}{rgb}{1,1,1}
\sethlcolor{LightBlue}
\definecolor{LightCyan}{rgb}{0.88,1,1}

\usepackage{xcolor}

\begin{document}
\bstctlcite{IEEEexample:BSTcontrol}
\title{\huge Sum-Rate Maximization of RSMA-based Aerial Communications with Energy Harvesting\\: A Reinforcement Learning Approach}
\author{Jaehyup Seong, \IEEEmembership{Student Member, IEEE}, Mesut Toka, \IEEEmembership{Member, IEEE},\\ and Wonjae Shin, \IEEEmembership{Senior Member, IEEE}\vspace{-5mm}

    }

\maketitle
\begin{abstract}
In this letter, we investigate a joint power and beamforming design problem for rate-splitting multiple access (RSMA)-based aerial communications with energy harvesting, where a self-sustainable aerial base station serves multiple users by utilizing the harvested energy. 
Considering maximizing the sum-rate from the long-term perspective, we utilize a deep reinforcement learning (DRL) approach, namely the soft actor-critic algorithm, to restrict the maximum transmission power at each time based on the stochastic property of the channel environment, harvested energy, and battery power information.
Moreover, for designing precoders and power allocation among all the private/common streams of the RSMA, we employ sequential least squares programming (SLSQP) using the Han–Powell quasi-Newton method to maximize the sum-rate for the given transmission power via DRL.
Numerical results show the superiority of the proposed scheme over several baseline methods in terms of the average sum-rate performance.
\end{abstract}
\vspace{-2mm}

\IEEEpeerreviewmaketitle

\vspace{-2mm}

\section{Introduction}

Unmanned aerial vehicle (UAV) communications have drawn a great deal of attention in the last few years, both in  academia and industry \cite{zeng2016wireless}. UAVs can be served not only as users but also as flying or aerial base stations (ABSs).
The deployment of ABSs enables supporting  ubiquitous connectivity, particularly in disaster and rural areas, and also provides high data rates in urban and suburban areas with favorable line-of-sight (LOS) propagation conditions. Thanks to their ability to extend network coverage and ensure high data rates, ABSs have emerged as one of the key enabling technologies for 5G networks and beyond \cite{liu2020opportunistic}. However, ABSs suffer from interference much more than terrestrial base stations (BSs)  due to the large and moving coverage areas while utilizing multi-antenna technologies to serve multiple users over the same frequency/time resource. Moreover, there are rapid channel variations due to the relative movement of ABSs with respect to ground users, \hl{thus acquiring perfect instantaneous channel state information (CSI) at the transmitter (CSIT) or the receiver (CSIR) becomes a challenging issue.}


To overcome the limitation of imperfect CSI in practical multi-antenna systems,   rate-splitting multiple access (RSMA)  has been recognized as a promising interference management strategy in various networks and propagation conditions \cite{mao2018rate}. 
In \cite{clerckx2019rate}, it has been revealed that RSMA can embrace conventional multiple access techniques and thus outperform in the presence of perfect CSI. Furthermore, it has been shown in \cite{an2021rate} \hl{that RSMA can still outperform the conventional schemes in terms of sum-rate even in the absence of both accurate CSIT and CSIR.}   


Inspired by this, in \cite{lin2021supporting}, RSMA-enabled UAV was employed to maximize the system aggregate rate in satellite-aerial integrated networks. Moreover, the authors of \cite{jaafar2020downlink} and \cite{jaafar2020multiple} have investigated ABS networks using RSMA to optimize the UAV location and sum-rate jointly. \hl{However, the self-sustainability of ABS networks has not been considered.} In other words, with no consideration regarding realistic constraints of ABSs such as insufficient power supply, ABSs cannot serve users continuously due to their limited power. 
To tackle this issue, energy harvesting-aided ABSs have emerged as a key solution with the intent of prolonging flight lifetime. Indeed, the authors of \cite{morton2015solar} developed solar-powered UAVs and showed that solar energy can be harvested for over 300 $\%$ of the power required for flight. Thus, the remaining power from the flight can be used in communications. In practice, the authors in \cite{sun2019optimal} designed an optimal policy for maximizing the system throughput from the long-term perspective based on the orthogonal multiple access (OMA) in solar-powered ABS networks.
However, in \cite{sun2019optimal}, it was assumed that ABSs have perfect statistical CSI knowledge, and harvestable energy is determined according to ABSs' locations, resulting in a lack of reality. Moreover, since OMA has been employed, the frequency band cannot be effectively used, which in turn degrades the spectral efficiency.

\hl{Different from the existing works, we propose a novel deep reinforcement learning (DRL)-based power allocation framework in energy harvesting-enabled ABS networks with RSMA to maximize the average sum-rate. Furthermore, realistic constraints such as randomness of energy arrival, time-varying channels, imperfect CSI, and finite-sized batteries are considered. Moreover, it is assumed that ABSs cannot have any prior knowledge of future arrival energy and CSI.} 
The main contributions are as follows:\vspace{-1mm}
\begin{itemize}
	\item We propose a robust power allocation and precoder design framework that maximizes the sum-rate from the long-term perspective for an RSMA-based ABS network with a stochastic energy harvesting model. In order to allocate power efficiently in the real-world environment, we perform the optimal power control by DRL approach, named soft actor-critic (SAC) algorithm \cite{haarnoja2018soft}.
	\item By taking the non-convexity of the precoding problem into account, we derive near-optimum precoding vectors in an iterative manner using the sequential least squares programming (SLSQP) algorithm\cite{kraft1988software}. \hl{In order to lighten the computational complexity from a highly accurate second-order approximation of SLSQP, the minimum mean square error (MMSE) method is used to design the normalized precoding vectors for private messages.} The rest of the RSMA parameters are set by the SLSQP algorithm.
	\item Numerical results demonstrate that the optimal power allocation policy in  energy harvesting networks is more effective in terms of the sum-rate. Also, employing the RSMA in considered networks significantly improves the sum-rate when compared with conventional multiple access techniques for both perfect and imperfect CSI.  
\end{itemize}




\begin{figure*}[!t]
\begin{equation}\label{RcEq1}
\small
    {R_{{\sf c},k}^{(i)}} =\log_{2}
    \bigg(1+\frac{|({\hat{\mathbf{h}}_{k}}^{(i)})^{\sf H} {\mathbf{p}_{\sf c}^{(i)}}|^2}{\sum\limits_{j=1}^{K}|({\hat{\mathbf{h}}_{k}}^{(i)})^{\sf H}{\mathbf{p}_{j}^{(i)}}|^2 + \sum\limits_{j\in{\mathcal{L}}}\mathbb{E}[|(\mathbf{e}_{k}^{(i)})^{\sf H}{\mathbf{p}_{j}^{(i)}}|^2] + \sigma_n^2} \bigg).
    \end{equation}
    \vspace{-4mm}
\end{figure*}
\begin{figure*}[!t]
\begin{equation}
    \label{RkEq2}
    \small
    {R_{k}^{(i)}} =\log_{2}\bigg(1+\frac{|({\hat{\mathbf{h}}_{k}}^{(i)})^{\sf H}{\mathbf{p}_{k} ^{(i)}}|^2}{\sum\limits_{j=1,j\neq{k}}^{K}|({\hat{\mathbf{h}}_{k}}^{(i)})^{\sf H}{\mathbf{p}_{j}^{(i)}}|^2 + \sum\limits_{j\in{\mathcal{L}}}\mathbb{E}[|(\mathbf{e}_{k}^{(i)})^{\sf H}{\mathbf{p}_{j}^{(i)}}|^2] + \sigma_n^2} \bigg).
    \end{equation}
    \noindent\rule{\textwidth}{.5pt}
\end{figure*}


\section{System Model and Problem Formulation}

\begin{figure}[!t]
\centering
 		\includegraphics[width=0.65\linewidth]{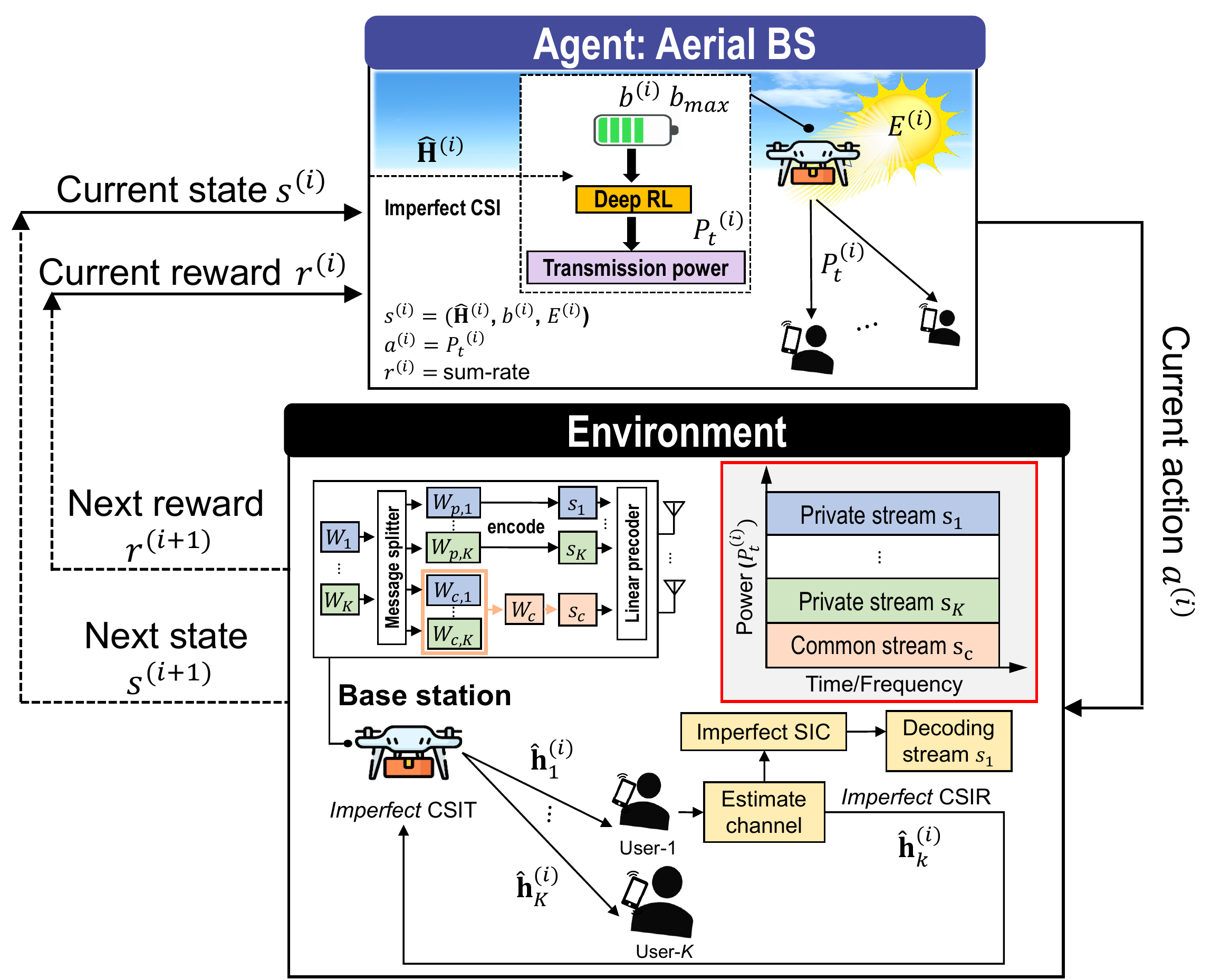}
 		\caption{System model of DRL based RSMA with energy harvesting ABS.}
    	\label{Fig1}\vspace{-4mm}
\end{figure}

We consider a \hl{multi-user multiple-input single-output (MU-MISO)} network as illustrated in Fig.~\ref{Fig1}, where an ABS simultaneously serves $K$ single-antenna users. 
To this end, the ABS harvests energy irregularly from renewable energy sources according to the environmental condition, followed by allocating the optimal total transmission power.
It then transmits desired signals to the users by using the allocated power. For a detailed explanation of the energy harvesting process for the ABS network, we denote a superscript $i$ as the time index. Firstly, the ABS with hybrid energy harvesting mechanism as in \cite{quyen2020optimizing} harvests energy $E^{(i)}$ stochastically from renewable energy sources (e.g., solar power and ambient RF radiation) with the energy harvesting probability ${p}_e$ at each time slot.
After the ABS replenishes its energy,  it broadcasts a linearly precoded signal to users using the total transmission power $P_{t}^{(i)}$ during transmission time $T$ by utilizing the remaining battery  $b^{(i)}$. The ABS then updates the battery status for the next time slot $b^{(i+1)}$ based on the amount of harvested energy $E^{(i)}$. However, the rechargeable battery has maximum energy storage, denoted by $b_{\sf max}$, due to its hardware constraint. Thus, the $b^{(i+1)}$ can be denoted as $\min\{{{b}^{(i)}-{T {P_{t}^{(i)}}} + {E^{(i)}},\, b_{{\sf max}}\}}$, and it is crucial for the ABS to allocate appropriate transmission power by considering the remaining battery, harvested energy, and channel quality of users. Since ABSs can provide a dominant LOS link with high probability, channels between the ABS and users, $\mathbf{h}_{k}\in\mathbb{C}^{N_t\times 1}$ are assumed to be exposed to Rician fading. Therefore, the signal received at the user $k$ can be expressed as $y_k = \mathbf{h}_k^{\sf H} \mathbf{x} + n$, where $\mathbf{x}\in \mathbb{C}^{N_t\times 1}$ represents the signal vector transmitted from the ABS, and $n\sim\mathcal{CN}(0,\sigma_n^2)$ denotes complex additive white Gaussian noise (AWGN). Assuming the imperfect CSIR due to the channel estimation error, the erroneous CSI vector can be expressed as $\mathbf{\hat{h}}_k = \mathbf{h}_k - \mathbf{e}_k \in \mathbb{C}^{N_t\times 1}$, where $\mathbf{e}_k\in \mathbb{C}^{{N_t}\times{1}}$ denotes the channel estimation error vector modeled as $\mathbf{e}_k$, and $\mathbf{h}_k$ denotes the actual CSI vector.  

Since we consider imperfect CSIR and CSIT, the concept of generalized mutual information can be used to determine the achievable rate for users \cite{an2021rate}. As in \cite{an2021rate}, the rate expressions of common and private messages for the $k$-th user at time slot $i$ can be formulated as (\ref{RcEq1}) and (\ref{RkEq2}) given at the top of this page, where $j \in \mathcal{L} \triangleq \{{\sf c},1,\cdots, K\}$. Here, ${\mathbf{p}_{\sf c}^{(i)}} \in \mathbb{C}^{{N_t}\times{1}}$ and ${\mathbf{p}_{k}^{(i)}} \in \mathbb{C}^{{N_t}\times{1}}$ denote the precoding vectors for the common and private messages, i.e.,
\begin{align}
    \label{Eq3}
   {\mathbf{p}_{\sf c}^{(i)}} = \sqrt{{P_{t}^{(i)}}{\mu_{\sf c}^{(i)}}}{\mathbf{w}_{\sf c}^{(i)}},
  ~~
    {\mathbf{p}_{k}^{(i)}} = \sqrt{{P_{t}^{(i)}}{\mu_{k}^{(i)}}}{\mathbf{w}_{k}^{(i)}},
\end{align}
where ${\mu_{\sf c}^{(i)}}$ $({\mu_{k}^{(i)}})$ and
${\mathbf{w}_{\sf c}^{(i)}}$ (${\mathbf{w}_{k}^{(i)}}$) $\in \mathbb{C}^{{N_t}\times{1}}$ denote the power ratios and normalized precoding vectors for common (private) messages, respectively. Since the power usage at time $i$ must not exceed the total power ${{P}_{t}^{(i)}}$, it follows ${\mu_{\sf c}^{(i)}} + \sum_{k=1}^{K} {\mu_{k}^{(i)}} = 1$.   $\mathbb{E}[|(\mathbf{e}_{k}^{(i)})^{\sf H}{\mathbf{p}_{j}^{(i)}}|^2]$ can be expressed as $(\mathbf{p}_{j}^{(i)})^{\sf H}{\mathbf{\Phi}_{k}^{(i)}}{\mathbf{p}_{j}^{(i)}}$ with the covariance matrix of the estimation error vector $\mathbb{E}[{\mathbf{e}_{k}^{(i)}} (\mathbf{e}_{k}^{(i)})^{\sf H}] = {\mathbf{\Phi}_{k}^{(i)}}\in \mathbb{C}^{{N_t}\times{N_t}}$.
Also, ${R_{\sf c}^{(i)}}=\underset{k}{\min}~{R_{{\sf c},k}^{(i)}}$ should be satisfied because the common message should be decoded by all users.

Therefore, the optimization problem  to maximize the total sum-rate during a total time  $T_o$ can be formulated as: 
\begin{align}\label{SREq4}
\max_{{\mathbf{p}_{\sf c}^{(i)}}, {\mathbf{p}_{1}^{(i)}}, \cdots, {\mathbf{p}_{K}^{(i)}}, {P_{t}^{(i)}}}{\sum_{i=0}^{T_o}R_{\sf sum}^{(i)}}
\end{align}\setcounter{equation}{3}\vspace{-4mm}
\begin{subequations}\label{condition}
\begin{align}
\mathrm{s.t.}\,\,\,\,\,\,
&{b}^{(i)}=\min{\{{b^{(i-1)}}-{P_{t}^{(i-1)}}T + {{E}^{(i-1)}},\, b_{{\sf max}}\}},\\
&{P_{t}^{(i)}} \leq {\frac{{b}^{(i)}}{T}},\\
&\sum_{j \in \mathcal{L}} \Vert {\mathbf{p}_{j}^{(i)}} \Vert ^2 \leq {P_{t}^{(i)}}, \ {R_{{\sf c},k}^{(i)}}\,{\geq {R_{\sf c}^{(i)}}},
\end{align}
\end{subequations}
where $R_{\sf sum}^{(i)} = R_{\sf c}^{(i)} + \sum_{k=1}^{K}{R_{k}^{(i)}}$. It is worth pointing out that the key problem addressed in this letter involves maximizing  the sum-rate from the long-term perspective, i.e., total sum-rate, in a self-sustainable network, which is a practical scenario for the ABS that operates continuously. Thus, the ultimate goal is to maximize the $\sum_{i=0}^{T_o}R_{\sf sum}^{(i)}$, not $R_{\sf sum}^{(i)}$ only.

\section{Long-term Achievable Sum-Rate Maximization based on DRL approach}

\begin{figure}[!t]
\centering
 		\includegraphics[width=0.65\linewidth]{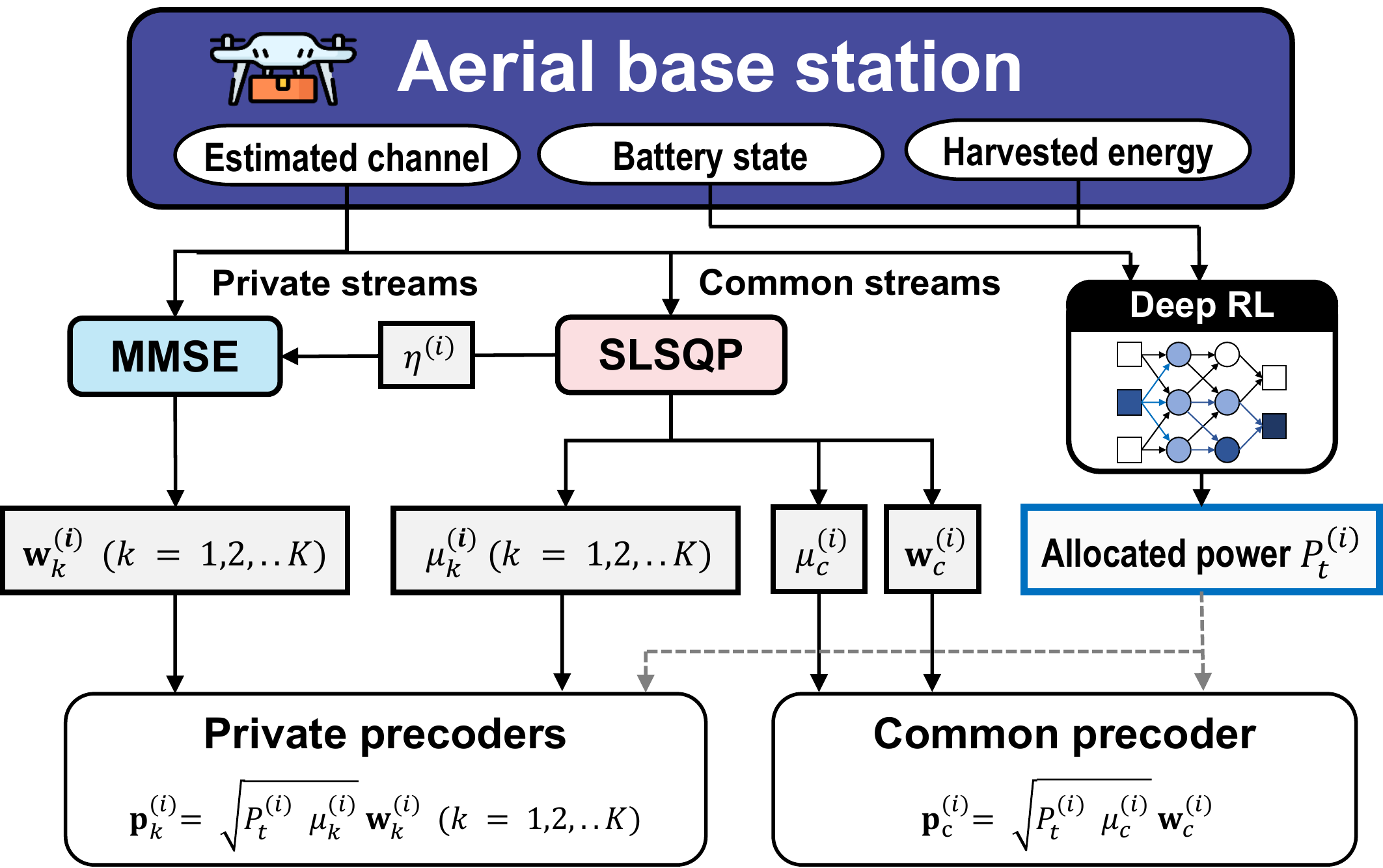}
 		\caption{\hl{Flow diagram of the proposed approach at time slot $i$.}}
     	\label{Fig2}
     	\vspace{-4mm}
\end{figure}

In this section, we first reformulate our problem into the Markov Decision Process (MDP). By doing so, the optimal instantaneous transmission power is determined  in each time slot using the SAC algorithm to maximize the total sum-rate over the total time  $T_{o}$.
\hl{Once the transmission power is allocated, the optimized RSMA precoder maximizing the instantaneous sum-rate for each time slot is obtained using the SLSQP algorithm and MMSE method.} Fig. \ref{Fig2} illustrates the flow diagram of the proposed approach for each time slot.

\subsection{Formulation of the Markov Decision Process}
When energy is harvested by the ABS, the remaining battery at time slot $i$ can be calculated as $\min\{{{b}^{(i-1)}-{T {P_{t}^{(i-1)}}} + {E^{(i-1)}},\, b_{{\sf max}}\}}$, where $T$ denotes the transmission time at each time slot.  \hl{The variables at the $i$-th time step respectively exist in each space, that is, remaining battery space ($\mathcal{B}$), harvested energy space ($\mathcal{E}$), and transmission power space ($\mathcal{P}_t$) such that ${{b}^{(i)}} \in \mathcal{B}$, ${{E}^{(i)}} \in \mathcal{E}$, and ${{P}_{t}^{(i)}} \in [0, {\frac{{b}^{(i)}}{T}}] \cap \mathcal{P}_{t}$.} The ABS is assumed to know imperfect CSIT related to the users at time slot $i$ as ${\hat{\mathbf{H}}}^{(i)} \in \mathbb{C}^{{N_t}\times{K}}$, where ${\hat{\mathbf{H}}^{(i)}} \in \mathcal{H}$ and $\mathcal{H}$ is continuous space.
We formulate the problem using the MDP, and hence, we define a tuple 
($\mathcal{S}, \mathcal{A},\mathbb{P}, r,\gamma$). Here, $\mathcal{S}$ denotes the state-space, $\mathcal{A}$ denotes the action space, and $\mathbb{P}$: $\mathcal{S} \times \mathcal{A} \times \mathcal{R} \rightarrow \mathcal{S}$ represents the state transition probability function of the next state information for the given state information and action. Additionally, $r$: $\mathcal{S} \times \mathcal{A} \rightarrow \mathcal{R}$ denotes the reward function, and $\gamma \in (0, 1)$ denotes the discount factor. The state information in the ABS is expressed as ${s}^{(i)} = ({E}^{(i)}, {\hat{\mathbf{H}}}^{(i)}, {b^{(i)}}) \in \mathcal{S}$, where $\mathcal{S} = \mathcal{E} \times \mathcal{H} \times \mathcal{B}$ denotes the state-space that is continuous. 
Meanwhile, the action-state information at the ABS is expressed as ${{a}^{(i)}} = {P_{t}}^{(i)} \in \mathcal{A}$, where $\mathcal{A}$ represents the action-space, which is also continuous. 
When the ABS uses the allocated transmission power ${P_{t}^{(i)}}$ into $s^{(i)}$, the instantaneous sum-rate $R_{\sf sum}^{(i)}$ can be re-expressed as the reward function $R({{s}^{(i)}}, {P_{t}^{(i)}})$.
In summary, we denote the state ${s}^{(i)} = ({E}^{(i)}, {\mathbf{\hat{H}}}^{(i)}, {b}^{(i)})$, action ${a}^{(i)} = {P_{t}^{(i)}}$, and reward function ${r}^{(i)} =R({{s}^{(i)}}, {P_{t}^{(i)}})$ at time slot $i$ to allocate the appropriate power to the ABS in the considered system.

\subsection{Optimized Power Allocation Policy Using SAC Algorithm}
As mentioned in Section \uppercase\expandafter{\romannumeral2}, our aim is to maximize the total sum-rate over the total time $T_o$. Here, since  $T_o$ can vary in the real-world environment, it is reasonable to consider the expected value of the total sum-rate to $T_o$ that can be defined as:
\begin{align}\label{ValFuncEq5}
    V(s) = \mathbb{E}_{\pi}
    \left[{\mathbb{E}_{T_o}}{\bigg[\sum_{i=0}^{T_o}
    {R({{s}^{(i)}},{P_{t}^{(i)}})}\bigg\vert\pi\bigg]}\right].
    \end{align}
 If $T_o$ is assumed to follow a geometric distribution with a mean of $1/(1-\gamma)$, ${\mathbb{E}_{T_o}}{[\sum_{i=0}^{T_o}{R({{s}^{(i)}},{P_{t}^{(i)}})}]}$ can be reformulated as $\sum_{T_o=0}^{\infty}((1-\gamma)\gamma^{T_0}\sum_{i=0}^{T_o}{R({{s}^{(i)}},{P_{t}^{(i)}}))}$. Then, it can be calculated as $\sum_{i=0}^{\infty}\gamma^{(i)}{R({{s}^{(i)}},{P_{t}^{(i)}})}$, and hence (\ref{ValFuncEq5}) can be interpreted as a discounted sum of the instantaneous sum-rate on the infinite time with a discount factor $\gamma$ \cite{kim2021shallow}. Thereby, (\ref{ValFuncEq5}) can be represented as:
\begin{align}\label{ValFuncEq6}
    V(s) = 
    {\mathbb{E}_{\pi}}{\bigg[\sum_{i=0}^{\infty}
    {{\gamma^{(i)}}R({{s}^{(i)}},{P_{t}^{(i)}})\bigg\vert\pi}\bigg]},
    \end{align}
where $V$ is referred to as the value function to measure the value of the state.

In the standard DRL, the optimal policy ${\pi}^*$ maximizes only the expected discounted reward sum in (\ref{ValFuncEq6}) in the environment. However, in the SAC, entropy regularization, which encourages both broader exploration and observation of several near-optimum in a continuous space, also needs to be considered. Therefore, $\pi^*$ can be represented as:
\begin{align}\label{Eq7}
{\pi}^* = \argmax_{\pi \in \Pi}\,\,
    {\mathbb{E}_{\pi}}{\bigg[\sum_{i=0}^{\infty}
    {{\gamma^{(i)}}R({{s}^{(i)}},{P_{t}^{(i)}})
    +\alpha H(\pi(\cdot\vert {{s}^{(i)}})) \bigg\vert\pi
    \bigg]}}   \nonumber
\end{align}\setcounter{equation}{6}\vspace{-4mm}
\begin{subequations}
\begin{align}
\mathrm{s.t.~~~~(4a), \,\,\,(4b)},
\end{align}
\end{subequations}
where $\Pi$ denotes the set of feasible policies in the MDP, $H(\pi(\cdot\vert {{s}^{(i)}}))$ denotes the entropy of the policy at state ${{s}^{(i)}}$, and temperature parameter $\alpha$ denotes a parameter that controls the trade-off between exploitation and exploration. 
The total transmission power ${P_{t}^{(i)}}$ is determined by the optimal policy $\pi^{*}$, where ${\pi^{*}}({{s}^{(i)}})$ = ${P_{t}^{(i)}}$.
By following the optimal policy ${\pi}^*$ for the power allocation, the optimal value function ${V}^*$, which denotes a measure of the long-term maximum achievable sum-rate of the state $s^{(i)}$, can be obtained. Policy evaluation and improvement are accomplished by training neural networks using stochastic gradient descent to determine ${P_{t}^{(i)}}$. Detailed information about this process can be found elsewhere \cite{haarnoja2018soft}.

\subsection{\hl{RSMA Precoder Based on MMSE and SLSQP}}
After the total power is allocated at the ABS by the above power allocation policy, that is, $\pi^{*}$ at time slot $i$, the messages of each user are split into common and private messages according to the RSMA strategy. Subsequently, the normalized precoding vector denoted by $\mathbf{w}_j^{(i)}$ is determined to design precoders according to (\ref{Eq3}). \hl{First, to address the inter-user interference with lower computational complexity, the normalized precoding vectors of the private messages are derived by the MMSE method, which can be formulated as:}
\begin{align}
     \label{Eq8}
     &{\mathbf{Z}^{(i)}}= {\mathbf{\hat{H}}^{(i)}}\bigg(({\mathbf{\hat{H}}^{(i)}})^{\sf H}{\mathbf{\hat{H}}^{(i)}} + \eta^{(i)} \frac{N_{t}}{P_t^{(i)}}(\sigma_{n}^2 + \frac{P_t^{(i)}}{N_{t}} \sigma_{e}^2)\mathbf{I}_{K}\bigg)^{-1}, \nonumber
    \\ &{\mathbf{Z}^{(i)}}= [{{\mathbf{z}}_{1}^{(i)}}, \dots, {{\mathbf{z}}_{K}^{(i)}}], \,\,\,
     {{\mathbf{w}}_{k}^{(i)}} = \frac{{{\mathbf{z}}_{k}^{(i)}}}{\Vert {{\mathbf{z}}_{k}^{(i)}} \Vert},
   \end{align}
\hl{where $\eta^{(i)}$ is an optimization variable to reflect the effect of the common stream caused by the imperfect SIC.}

\hl{Owing to the non-convexity of maximizing the instantaneous sum-rate for each time step $i$, denoted as $R_{\sf sum}^{(i)}$, it is still difficult to directly obtain the optimal values of $\eta^{(i)}$, ${\mu_{\sf c}^{(i)}}$, ${\mu_{k}^{(i)}}$, and ${\mathbf{w}_{\sf c}^{(i)}}$.} To address this issue, we use the SLSQP algorithm, which is used to solve non-linear (NL) problems in an iterative manner. The SLSQP algorithm is based on the sequential quadratic programming (SQP) algorithm, which transforms the constrained NL problem into a quadratic sub-problem by the second-order Taylor series expansion and updates the sub-problem to approximate the original NL problem iteratively. There exist various approaches based on the SQP algorithm, however, the SLSQP algorithm is the most advanced one, thanks to its low computational complexity, superior performance, and super-linear and global convergence speeds \cite{kraft1988software}. The procedure of the SLSQP algorithm to derive optimal values can be summarized as follows, where the subscript $\tau$ denotes the step-index in the SLSQP algorithm. It is worth emphasizing that all procedures conducted are included at each time slot $i$. \hl{The initial iteration point is composed of initialized values of $\eta^{(i)}$, ${\mu_{k}^{(i)}}$, ${\mu_{\sf c}^{(i)}}$, and the real and imaginary parts of ${\mathbf{w}_{\sf c}^{(i)}}$, denoted as $\mathbf{x}_{0}^{(i)} \in \mathbb{R}^{(K+2+2{N_t})\times{1}}$.} 
\begin{itemize}
	\item \textbf{Step 1:} Construct the quadratic sub-problem of (\ref{SREq4}) using the second-order Taylor expansion, with the initial point and Hessian matrix of the Lagrangian for (\ref{SREq4}), that is, \\ \hl{$\mathbf{W}_{0}^{(i)} \in \mathbb{R}^{(K+2+2{N_t})\times(K+2+2{N_t})}$.} Due to the complexity in calculating the Hessian matrix, the Wilson-Han-Powell method \cite{kraft1988software} is adopted to replace $\mathbf{W}_{0}^{(i)}$ with the positive definite matrix \hl{$\mathbf{A}_{0}^{(i)} \in \mathbb{R}^{(K+2+2{N_t})\times(K+2+2{N_t})}$} under suitable assumptions.
	\item \textbf{Step 2:} Solve the constructed quadratic sub-problem and test whether the termination condition is satisfied. If so, the current solution $\mathbf{x}_{\tau}^{(i)}$ is regarded as the solution to the original problem (\ref{SREq4}), and the iteration is terminated.   
	\item \textbf{Step 3:} Otherwise, use the line search method by adopting the $L_1$-norm as the loss function to calculate the search step length $\alpha_{\tau}^{(i)}$ in the current direction.  
	\item \textbf{Step 4:} Update the symmetric definite matrix $\mathbf{A}_{\tau}^{(i)}$ using the Han–Powell quasi-Newton method with a BFGS update \cite{kraft1988software} and update the iteration point $\mathbf{x}_{\tau}^{(i)}$ by using the search step length $\alpha_{\tau}^{(i)}$. Then, reconstruct the quadratic sub-problem and revisit \textbf{Step 2} for the next iteration.  
\end{itemize}

To apply the SLSQP algorithm to the defined problem by the aforementioned steps, we only need to provide objective, constraint functions, and the range of the optimization variable. Therefore, we set the objective to maximize $R_{\sf sum}^{(i)}$ with constraint functions (\ref{SREq4}c) for each time slot $i$, while setting the ranges of the optimization variables as $0 \leq \eta^{(i)}$, $0 \leq {\mu_{\sf c}^{(i)}}, {\mu_{k}^{(i)}} \leq 1$, and ${\Vert{{\mathbf{w}}_{\sf c}^{(i)}}\Vert^2}{\leq 1}$.

\begin{table}[!t]\renewcommand{\arraystretch}{1.35} 
\caption{{Selected  parameter settings for learning}}
\label{Table1}
\centering
\begin{tabular}{|c||c|c||c|}
\hlineB{3}

\textbf{number of learning process} & {$10$} & \textbf{temperature} & {$0.5$}\\
\hhline{|----|}
\textbf{replay buffer size} & {$10^6$} & \textbf{batch size} & {$1024$}\\
\hhline{|----|}
\textbf{actor learning rate} & {$10^{-3}$} & \textbf{critic learning rate} & {$10^{-3}$}\\
\hhline{|----|}
 \hlineB{3}
\end{tabular}\vspace{-3mm}
\end{table}

\section{PERFORMANCE EVALUATION}

\hl{In this section, we present numerical results for two-user MISO networks as an example, where ABS is equipped with two antennas ($K=2, N_{t}=2$).}
It is assumed that each element of $\mathbf{h}_k$ follows independent and identically distributed (i.i.d.) Rician fading such that $\mathbf{h}_k\sim\mathcal{CN} (s,2\sigma_h^2)$, where a Rician shape parameter $\kappa=\frac{s^2}{2\sigma_h^2}$ and scale parameter $\Omega=s^2+2\sigma_h^2$ is set as 3 and 1, respectively.\footnote{$\mathbf{h}_k=\mathbf{x}_k$+$j\mathbf{y}_k$ where $\mathbf{x}_k\sim\mathcal{N}(s,\sigma_h^2)$ and $\mathbf{y}_k\sim\mathcal{N}(0,\sigma_h^2)$; $s^2$ and $2\sigma_h^2$ are the average power of LOS and non-LOS components, respectively.}
$\mathbf{e}_k$ is independent of $\mathbf{h}_k$ and follows i.i.d. complex Gaussian distribution such that $\mathbf{e}_k\sim\mathcal{CN}(\mathbf{0},\sigma_{e,k}^2\mathbf{I})$, where each estimation error is assumed to have the same variance as $\sigma_e^2=0.1$.
Therefore, $\hat{\mathbf{h}}_k$ can be represented as $\hat{\mathbf{h}}_k\sim$ $\mathcal{CN}(s,2\sigma_h^2-0.1)$. 
The variance of AWGN $\sigma_n^2$ is fixed as 1.
Besides, we assume the energy harvesting probability as ${p}_e=0.5$ with the Bernoulli process, and the transmission time $T$ is set as $1$ so that the consumed energy ${P_{t}}T$ can be treated as ${P_{t}}$. 
As a performance metric, we measure the average sum-rate obtained by averaging the results of 10 learning processes, where the result of each learning process is an average sum-rate of 1000 time steps.
The remaining learning parameters are provided in Table \ref{Table1}.

\begin{figure}[!t]
\centering \vspace{-2mm}
 		\includegraphics[width=0.65\linewidth]{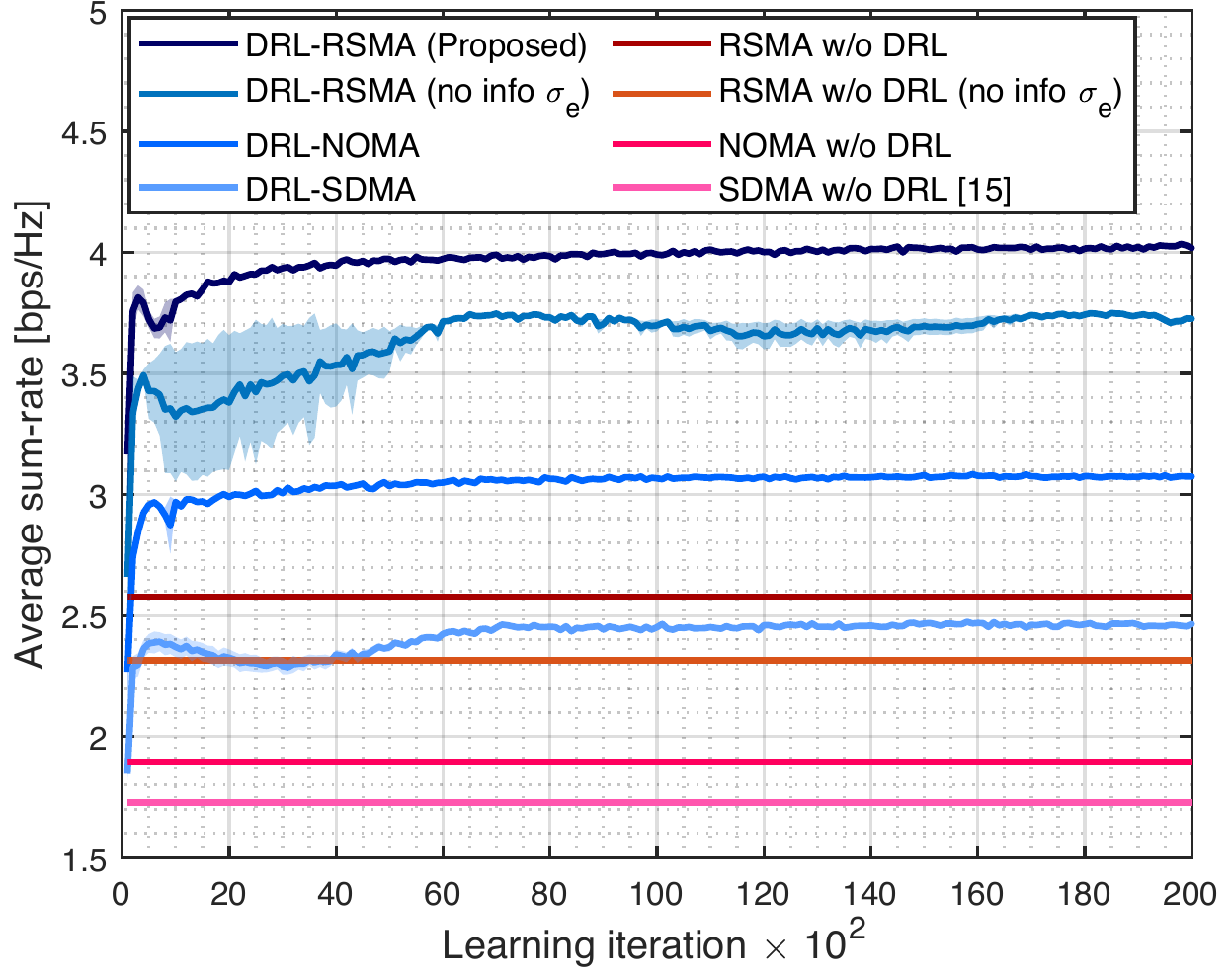}\vspace{-3mm}
 		\caption{
 		\hl{Comparisons of the proposed scheme (DRL-RSMA) with benchmark schemes versus learning iteration.}}
     	\label{Fig3}\vspace{-3mm}    
\end{figure}

\begin{figure}[!t]
\centering
 		\includegraphics[width=0.65\linewidth]{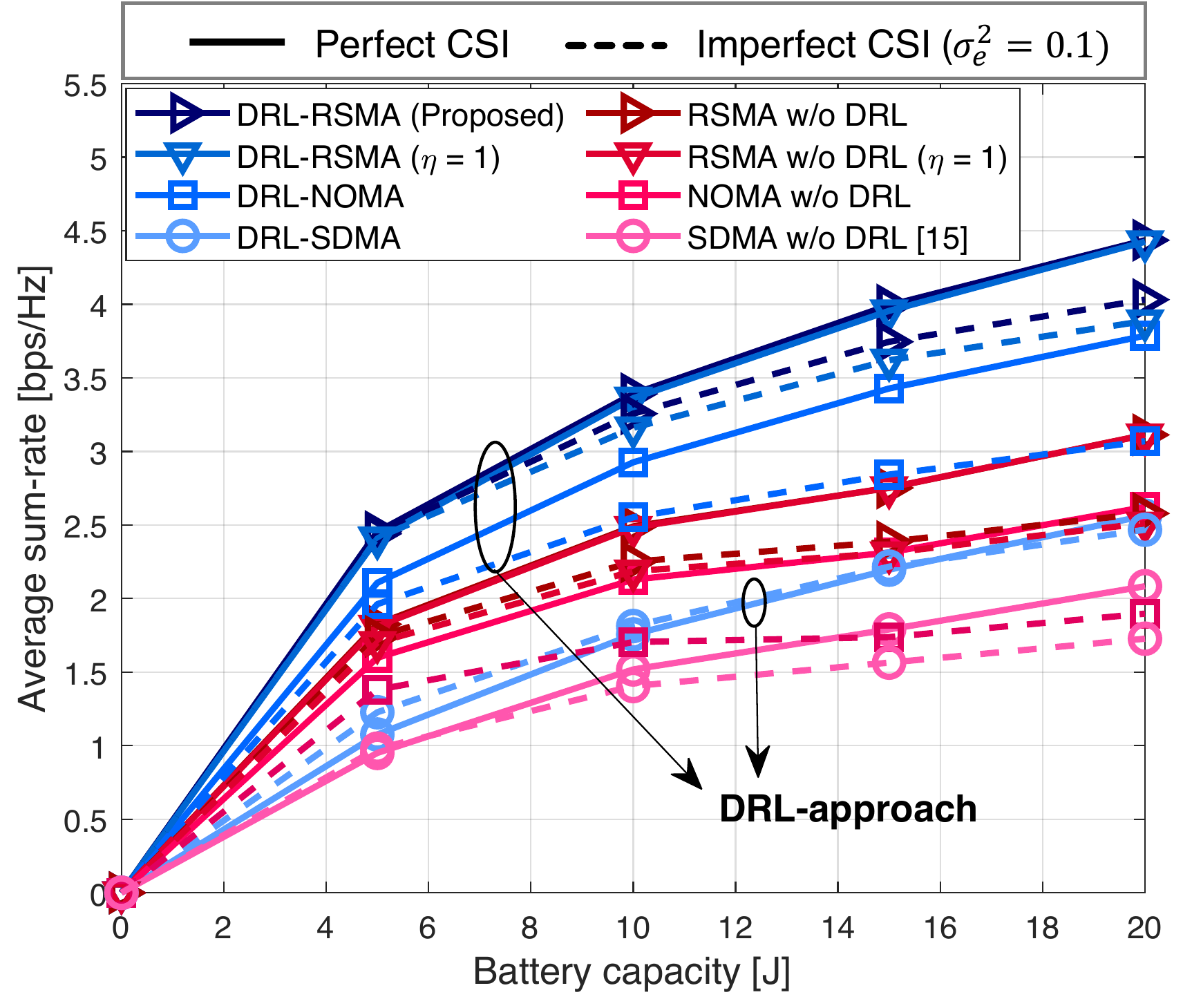}\vspace{-3mm}
 		\caption{
 		\hl{Average sum-rate comparisons of the proposed scheme (DRL-RSMA) with benchmark schemes versus battery capacity.}}
     	\label{Fig4}\vspace{-3mm}    
\end{figure}

In Fig. \ref{Fig3}, the average sum-rate performance of the proposed scheme (DRL-RSMA) versus learning iteration is presented, while both the battery capacity $b_{{\sf max}}$ and the maximum harvested energy $E_{{\sf max}}$ are set to 20 J. 
Here, the proposed DRL-RSMA is compared with DRL-RSMA (no-info $\sigma_e$) scheme, the non-orthogonal multiple access (NOMA) scheme, and the \cite{spencer2004zero} based spatial division multiple access (SDMA) scheme.
The no-info $\sigma_e$ indicates that the ABS does not have or ignore the statistical information of the channel estimation error, i.e., $\mathbf{p}_{j}^H\mathbf{\Phi}_{k}\mathbf{p}_{j}=\bf{0}$; thus, the ABS designs the beamforming by considering $\hat{\mathbf{h}}_k$ as a perfect channel estimate.
The rate performances of three different schemes relying on optimal power allocation with SAC are compared. \hl{The performance gap between the proposed DRL-RSMA and the DRL-NOMA / DRL-SDMA represents the superiority of the RSMA scheme.}
Moreover, the gap in performance and convergence speed between DRL-RSMA and DRL-RSMA (no-info $\sigma_e$) shows the impact of exploiting the second-order statistics of channel errors at the ABS.
Also, we compare the proposed scheme with greedy power allocation-based schemes that instantaneously use all of their harvested energy without saving it for future use. As illustrated, the performance gap between the RSMA w/o DRL and the other greedy schemes
are similar to the previous results. However, a significant performance gain from RSMA w/o DRL to DRL-RSMA can be observed from the effect of the adopted optimal power allocation policy. This result implies the importance of the power allocation policy for a self-sustainable network, and particularly, as illustrated, the proposed scheme rapidly converges with high stability.

In Fig. \ref{Fig4}, the average sum-rate of the proposed scheme is compared with the various kind of both optimal power allocation-based schemes and greedy power allocation-based schemes versus battery capacities for $\kappa=3$. Herein, DRL-RSMA ($\eta=1$) denotes the scheme that MMSE parameter $\eta$ is fixed as 1 in overall time steps. As can be observed from the figure, in the imperfect CSIT and CSIR, due to the effect of imperfect SIC according to the channel estimation error, the performance gap between the proposed scheme and DRL-RSMA ($\eta=1$) occur.\footnote{\hl{Our proposed scheme achieves nearly the same performance with low computational complexity compared to optimizing the entire parameters of the RSMA precoder with only SLSQP.}}
Moreover, the proposed scheme shows a higher performance than the other multiple access-based schemes even if perfect CSIT and CSIR are not available in all battery capacity regions. These results demonstrate the superiority of the proposed scheme in practical multi-antenna systems in which imperfect CSI usually arises due to the availability of noisy channel estimates at the receiver.



\section{Conclusion}
In this letter, we have investigated a robust design of RSMA-based MU-MISO aerial communications with stochastic energy harvesting models under imperfect CSIT and CSIR. To deal with the average sum-rate maximization problem, a DRL approach, namely the SAC algorithm and SLSQP algorithm have been jointly utilized. Simulation results have demonstrated that the proposed scheme achieves the best sum-rate performance compared with benchmark schemes in energy harvesting communications. 
\hl{Future directions include further optimizing the nature of ABSs as trajectory design or flight energy consumption for more realistic ABS networks.}

\bibliographystyle{IEEEtran}

\bibliography{jhseong_reff}

\begin{thebibliography}{10}
\providecommand{\url}[1]{#1}
\csname url@samestyle\endcsname
\providecommand{\newblock}{\relax}
\providecommand{\bibinfo}[2]{#2}
\providecommand{\BIBentrySTDinterwordspacing}{\spaceskip=0pt\relax}
\providecommand{\BIBentryALTinterwordstretchfactor}{4}
\providecommand{\BIBentryALTinterwordspacing}{\spaceskip=\fontdimen2\font plus
\BIBentryALTinterwordstretchfactor\fontdimen3\font minus
  \fontdimen4\font\relax}
\providecommand{\BIBforeignlanguage}[2]{{%
\expandafter\ifx\csname l@#1\endcsname\relax
\typeout{** WARNING: IEEEtran.bst: No hyphenation pattern has been}%
\typeout{** loaded for the language `#1'. Using the pattern for}%
\typeout{** the default language instead.}%
\else
\language=\csname l@#1\endcsname
\fi
#2}}
\providecommand{\BIBdecl}{\relax}
\BIBdecl

\bibitem{zeng2016wireless}
Y.~Zeng, R.~Zhang, and T.~J. Lim, ``Wireless communications with unmanned
  aerial vehicles: Opportunities and challenges,'' \emph{IEEE Commun. Mag.},
  vol.~54, no.~5, pp. 36--42, 2016.

\bibitem{liu2020opportunistic}
D.~Liu, Y.~Xu, J.~Wang, J.~Chen, K.~Yao, Q.~Wu, and A.~Anpalagan,
  ``Opportunistic {UAV} utilization in wireless networks: Motivations,
  applications, and challenges,'' \emph{IEEE Commun. Mag.}, vol.~58, no.~5, pp.
  62--68, 2020.

\bibitem{mao2018rate}
Y.~Mao, B.~Clerckx, and V.~O. Li, ``Rate-splitting multiple access for downlink
  communication systems: bridging, generalizing, and outperforming {SDMA} and
  {NOMA},'' \emph{EURASIP J. Wirel. Commun. Netw.}, vol. 2018, no.~1, pp.
  1--54, 2018.

\bibitem{clerckx2019rate}
B.~Clerckx, Y.~Mao, R.~Schober, and H.~V. Poor, ``Rate-splitting unifying
  {SDMA}, {OMA}, {NOMA}, and multicasting in {MISO} broadcast channel: A simple
  two-user rate analysis,'' \emph{IEEE Wireless Commun. Lett.}, vol.~9, no.~3,
  pp. 349--353, 2019.

\bibitem{an2021rate}
J.~An, O.~Dizdar, B.~Clerckx, and W.~Shin, ``Rate-splitting multiple access for
  multi-antenna broadcast channel with imperfect {CSIT} and {CSIR},''
  \emph{arXiv preprint arXiv:2102.08738}, 2021.

\bibitem{lin2021supporting}
Z.~Lin, M.~Lin, T.~De~Cola, J.-B. Wang, W.-P. Zhu, and J.~Cheng, ``Supporting
  {IoT} with rate-splitting multiple access in satellite and aerial-integrated
  networks,'' \emph{IEEE Internet of Things Journal}, vol.~8, no.~14, pp.
  11\,123--11\,134, 2021.

\bibitem{jaafar2020downlink}
W.~Jaafar, S.~Naser, S.~Muhaidat, P.~C. Sofotasios, and H.~Yanikomeroglu, ``On
  the downlink performance of {RSMA}-based {UAV} communications,'' \emph{IEEE
  Trans. Veh. Technol.}, vol.~69, no.~12, pp. 16\,258--16\,263, 2020.

\bibitem{jaafar2020multiple}
W.~Jaafar, S.~Naser, S.~Muhaidat, P.~C. Sofotasios, and H.~Yanikomeroglu,
  ``Multiple access in aerial networks: From orthogonal and non-orthogonal to
  rate-splitting,'' \emph{IEEE Open J. Veh. Technol.}, vol.~1, pp. 372--392,
  2020.

\bibitem{morton2015solar}
S.~Morton, R.~D'Sa, and N.~Papanikolopoulos, ``Solar powered {UAV}: Design and
  experiments,'' in \emph{2015 IEEE/RSJ international conference on intelligent
  robots and systems (IROS)}.\hskip 1em plus 0.5em minus 0.4em\relax IEEE,
  2015, pp. 2460--2466.

\bibitem{sun2019optimal}
Y.~Sun, D.~Xu, D.~W.~K. Ng, L.~Dai, and R.~Schober, ``Optimal {3D}-trajectory
  design and resource allocation for solar-powered {UAV} communication
  systems,'' \emph{IEEE Trans. on Commun.}, vol.~67, no.~6, pp. 4281--4298,
  2019.

\bibitem{haarnoja2018soft}
T.~Haarnoja, A.~Zhou, P.~Abbeel, and S.~Levine, ``Soft actor-critic: Off-policy
  maximum entropy deep reinforcement learning with a stochastic actor,'' in
  \emph{Proc. Int. Conf. Machine Learning}, 2018, pp. 1861--1870.

\bibitem{kraft1988software}
D.~Kraft, \emph{A software package for sequential quadratic programming}.\hskip
  1em plus 0.5em minus 0.4em\relax Wiss. Berichtswesen d. DFVLR Brunswick,
  Germany, 1988.

\bibitem{quyen2020optimizing}
T.~Quyen, C.~Nguyen, A.~Le, and M.~Nguyen, ``Optimizing hybrid energy
  harvesting mechanisms for {UAV}s,'' \emph{EAI Endorsed Transactions on Energy
  Web}, vol.~7, no.~30, 2020.

\bibitem{kim2021shallow}
H.~Kim, J.~Lee, W.~Shin, and H.~V. Poor, ``Shallow reinforcement learning for
  energy harvesting communications with imperfect channel knowledge,''
  \emph{IEEE J. Sel. Topics Signal Process.}, vol.~15, no.~5, pp. 1258--1271,
  2021.

\bibitem{spencer2004zero}
Q.~H. Spencer, A.~L. Swindlehurst, and M.~Haardt, ``Zero-forcing methods for
  downlink spatial multiplexing in multiuser {MIMO} channels,'' \emph{IEEE
  Trans. Signal Process.}, vol.~52, no.~2, pp. 461--471, 2004.

\end{thebibliography}
\end{document}